# Billiards in a gravitational field: A particle bouncing on a parabolic and right angle mirror

S. Masalovich

Heinz Maier-Leibniz Zentrum (MLZ), Technische Universität München,
Lichtenbergstr. 1, 85748 Garching, Germany
*Sergey.Masalovich@frm2.tum.de*

**Abstract**

We propose geometric tools that are suitable for studying the behavior of a billiard trajectory in a homogeneous force field. Two examples are considered: a vertical plane with an open top and with a parabolic or right angle boundary at the bottom. In either case, we obtain equations that describe the envelope of a trajectory. These equations are in good agreement with those found earlier with the use of other calculation methods. In addition, we present some new geometric properties of trajectories. We show that in the case of a parabolic boundary the sequence of the trajectory impact points can be easily constructed by multiple reflections of a single ellipse.

## 1. Introduction

Billiard systems are proved to be a useful tool in solving different problems in theoretical mechanics, topology, the number theory and, of course, in geometry [1,2]. In general, mathematical billiards describes a motion of a free point particle in a domain with a reflecting boundary. The reflection is assumed to be elastic, i.e. the angle of incidence equals the angle of reflection. In the absence of external forces, a particle moves along a straight line between two sequential impact points and thus we have here an idealization of an ordinary billiard-table game. In this virtual game a particle (a ball) keep going with no loss of energy and may bounce infinitely many times on the boundary. The study of the path that takes a particle (billiard trajectory) is the subject of mathematical billiards. In the present paper we shall consider the behavior of a particle bouncing on a parabolic mirror in the presence of a gravitational field. In this case, the billiard table represents a vertical plane with an open top. The only boundary is the parabolic mirror that opens upwards. Such a billiard system arises, for example, in designing of scientific instruments such as a neutron microscope with ultracold neutrons [3,4] or a storage cavity for trapped cold atoms [5,6]. In both cases, the energy of particles is so low that the gravity starts to play a significant role. The segments between two sequential impact points are no longer straight lines but parabolas (flight parabolas). Below we consider two types of billiard tables - a parabolic mirror and a right angle mirror. We shall develop geometric tools to tackle such problems, evaluate the long-term behavior of a billiard path and discuss the geometric aspects of the results.

## 2. Billiards with a parabolic boundary

Billiards with a parabolic boundary in the presence of the gravitational field has already been discussed and the unique properties of trajectories were reported [3-6]. These properties were studied with the use of algebraic calculations and this approach was shown to be a powerful method. We have earlier proposed a complementary geometric approach to such systems and illustrated it with the study of the focusing property of a parabolic mirror for trajectories that run through the focal point of the mirror [7]. We now extend that geometric approach and apply it to study the behavior of any trajectory. It will be shown that our results agree well with the results obtained in the study of atoms bouncing in a parabolic cavity [5], but also some new geometrical properties of trajectories will be presented.

Let us consider two-dimensional billiards in a XZ plane where the z-axis is aligned vertically. First, we write down a few well-known formulae, which will be referred to throughout the paper. A general equation of a parabolic boundary (mirror) in the frame of reference where the origin is located at the focus of the mirror is

$$z = \frac{x^2}{4F} - F. \tag{1}$$

Here the parabola opens up and its focal length is denoted by $F$.

We now take an arbitrary particle that runs through the origin with a velocity $v$ and at that point the trajectory makes an angle α with the vertical axis. By writing the equation of the trajectory in a parametric form

$$x(t) = v \cdot \sin\alpha \cdot t, \tag{2.1}$$

$$z(t) = v \cdot \cos\alpha \cdot t - \frac{gt^2}{2}, \tag{2.2}$$

and eliminating $t$ we obtain

$$z_1 = \frac{2g}{4v^2 \sin^2\alpha} \cdot x_1^2 - \frac{v^2 \sin^2\alpha}{2g}, \tag{3}$$

where $g$ is the acceleration due to gravity and $t$ is a flight time.
In Eq. (3) we have used the substitutions:

$$x_1 = x - \frac{v^2 \sin 2\alpha}{2g}, \tag{4.1}$$

$$z_1 = -z + \frac{v^2 \cos 2\alpha}{2g}. \tag{4.2}$$

The comparison of Eqs. (1) and (3) shows that the trajectory of the particle is a parabola with a focal length

$$f = \frac{v^2 \sin^2\alpha}{2g}. \tag{5}$$

From Eqs. (4.1) and (4.2) it follows that the focus of this parabola has coordinates:

$$x_f = \frac{v^2 \sin 2\alpha}{2g}, \qquad (6.1)$$

$$z_f = \frac{v^2 \cos 2\alpha}{2g}. \qquad (6.2)$$

Hence, the focus of any parabolic trajectory (irrespective of an initial angle) that runs through the origin with the same velocity $v$ lies on a circle of radius

$$R_0 = \frac{v^2}{2g} \qquad (7)$$

with the center at the origin [8]. Besides, all those trajectories (flight parabolas) have a common directrix represented as a horizontal straight line with z-coordinate $H$ defined by

$$z = H = z_f + 2f = \frac{v^2}{2g}. \qquad (8)$$

Consequently, the position of the directix depends only on the magnitude of the velocity vector as illustrated in Fig.1.

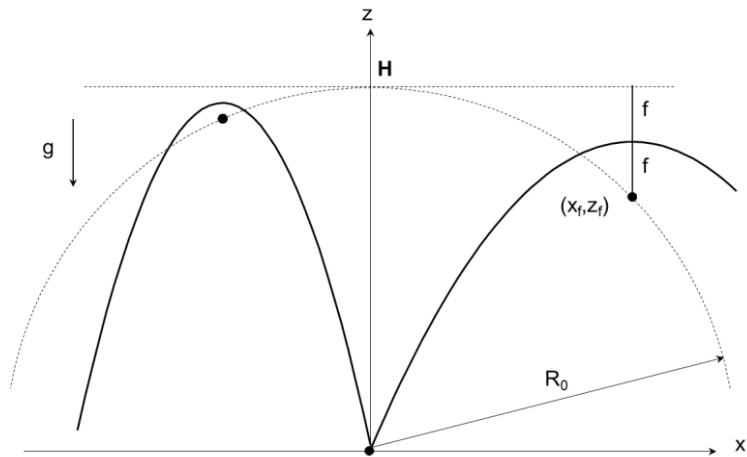

Fig.1. Two flight parabolas with the common directix $H$ and the foci laying on the common circle of radius $R_0$.

Eq. (7) was already used in our earlier paper [7] where we studied the focusing property of a parabolic mirror. It was shown that if a particle passed through the focal point of a mirror, then, after a specular reflection by that mirror, the particle would again run through this focal point. It holds true for all subsequent reflections. Moreover, the sequential parabolic arcs (flight parabolas) converge to the vertical line through the focus of the mirror as the number of reflections increases.

We now consider a general case when a particle does not pass through the focal point. We are to show that also in this case the foci of all sequential parabolic segments lie on a common circle. Let O be the origin (focus of the mirror) and A denotes the point where the collision occurs (see Fig.2).

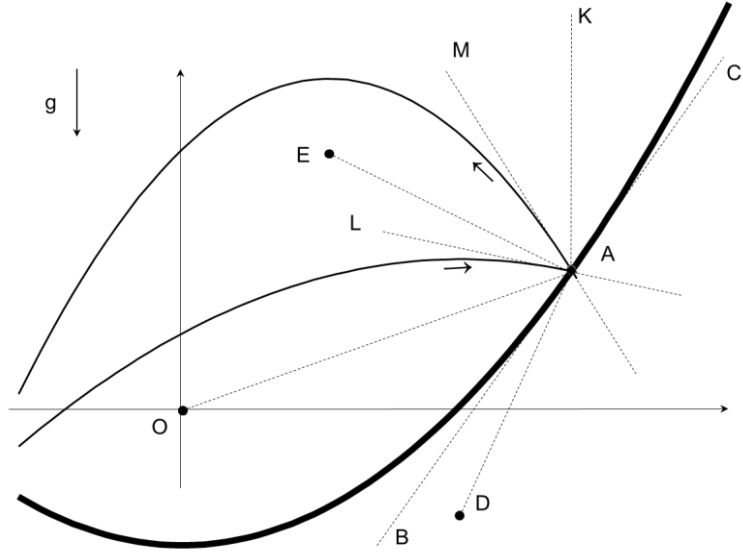

Fig.2. Reflection of a particle by a parabolic mirror. Here and in what follows the mirror is shown as an extra thick line.

The lines and points in Fig.2 are lettered as follows: AK – the vertical line, BC – the tangent to the mirror surface at A, AL and AM – the tangents to the parabolic arcs of an incoming and outgoing particle at A, and D and E– the focal points of those parabolic arcs. Because of a specular reflection and by the properties of a parabola, we have: ∠LAB = ∠MAC, ∠BAO = ∠CAK, ∠EAM = ∠MAK, and ∠DAL = ∠LAK. Then, by straightforward calculations, we find that

$$\angle EAO = \angle DAO. \qquad (9.1)$$

Since E and D are the focal points of the two parabolic paths of a particle that goes through A with the same velocity, then (see the deduction of Eq. (7))

$$DA = EA. \qquad (9.2)$$

Eqs. (9.1) and (9.2) imply that the points E and D are symmetric about the line OA and thus at an equal distance from O. Since this statement is true for any two successive parabolic segments of the same trajectory, we conclude that the foci of all flight parabolas of a particle bouncing on a parabolic mirror lie on a common circle, which we call a focal circle. The center of the focal circle lies at the focus of the mirror. Moreover, making use of the energy conservation law, it is easy to verify that all subsequent parabolic arcs have a common directrix. In the frame of reference with the origin at the focus of the mirror the z-coordinate of the directix is given by the equation

$$H = -F + \frac{v_F^2}{2g}, \qquad (10)$$

where $v_F$ is the magnitude of the particle velocity at the level $z = -F$ (the horizontal line through the vertex of the mirror). It follows from Eq. (10) that $H$ may be positive or negative depending on the velocity value.

We first consider the case $H > 0$. Fig.3 shows three trajectories of a particle with the same velocity vector at the level $z = 0$, but with different x-coordinates.

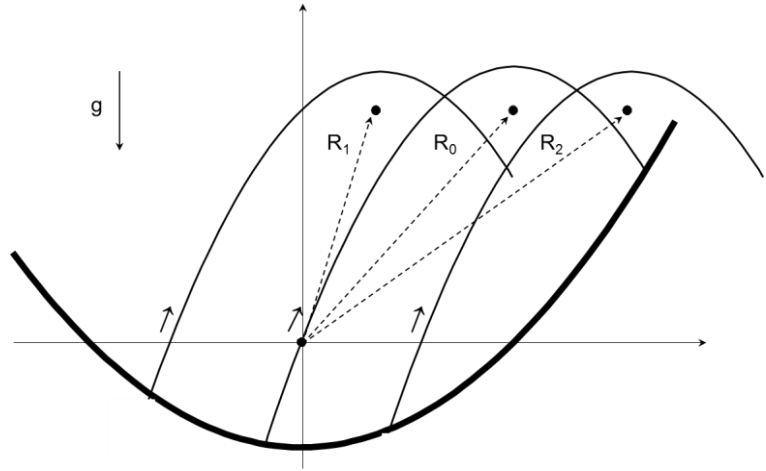

Fig.3. Three different trajectories of a particle with a common directix ($H > 0$).

It can clearly be seen that if initially a trajectory goes above the focus of the mirror then the radius $R_1$ of a focal circle is less than $R_0$. On the other hand, if initially a trajectory goes below the focus of the mirror, then the radius $R_2$ of a focal circle is greater than $R_0$. Therefore, for a particle that has at the level $z = 0$ a velocity $v$ and $x \neq 0$, one can get either $R > R_0$ or $R < R_0$ depending on an initial angle and x-coordinate.

If a particle has a very low energy such that it can never reach the level $z = 0$, then obviously its trajectory always lies below the focus of the mirror and the z-coordinate of the directix $H < 0$. Under these conditions we only have $R > |H|$, which follows from the energy conservation law.

Consequently, we conclude that if $R > |H|$ ($H > 0$ or $H < 0$) then a particle trajectory will always cross the z-axis below the focus of the mirror, and if $R < |H|$ ($H > 0$) then always above the focus of the mirror while bouncing on the mirror. Hence, a trajectory is bounded in some definite region of the parabolic table and we are interested to find the boundaries of this region. With this aim in mind, we recall that for any given bouncing particle two parameters, $R$ and $H$, remain constant and they determine the geometry of an area accessible to the particle. We shall consider a boundary of that area to be an envelope of the family of sequential parabolic arcs that make up the infinite trajectory of a bouncing particle. The typical flight parabola of this family is shown in Fig.4 where its focus is marked by $O'$ and the focal length is denoted by $f$. The radius-vector $R$ to the focus of the flight parabola makes an angle α with the vertical direction.

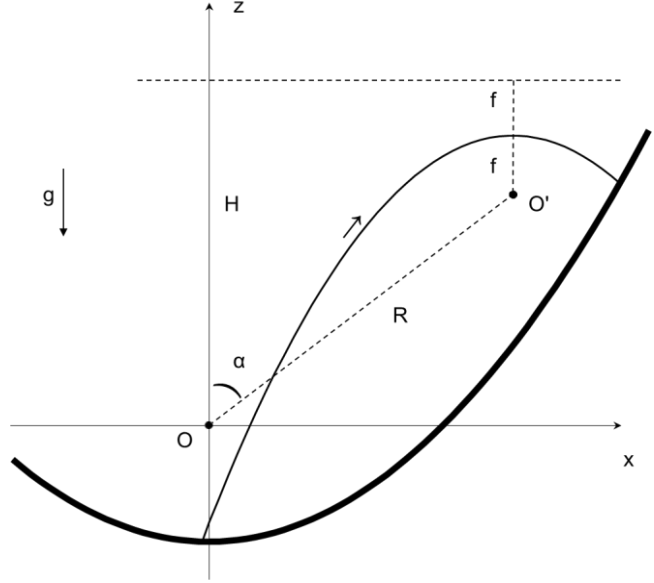

Fig.4. For a bouncing particle the parameters $R$ and $H$ remain constant, but the angle $\alpha$ changes

The equation of this parabola in the frame of reference with the origin at $O'$ reads

$$-z' = \frac{x'^2}{4f} - f. \qquad (11)$$

The substitutions

$$x = R\sin\alpha + x', \qquad (12.1)$$
$$z = R\cos\alpha + z', \qquad (12.2)$$

and the obvious relation

$$f = \frac{1}{2}(H - R\cos\alpha) \qquad (13)$$

allow us to rewrite Eq. (11) in the frame of reference with the origin at $O$:

$$F(x, z, \alpha, R, H) \equiv x^2 + 2zR + R^2 - H^2 - 2R(x\sin\alpha + z\cos\alpha) = 0. \qquad (14)$$

Obviously, Eq. (14) describes the family of all parabolic segments with given $R$ and $H$. The parameter $\alpha$ is a variable parameter within the family and determines any particular segment. Hence, the equation of the envelope can be obtained by solving the system [9]

$$F(x, z, \alpha, R, H) = 0, \qquad (15.1)$$
$$\partial F(x, z, \alpha, R, H)/\partial \alpha = 0. \qquad (15.2)$$

From Eq. (15.2) we find

$$\partial F/\partial \alpha = -R(x\cos\alpha - z\sin\alpha) = 0 \qquad (16)$$

and hence

$$\tan \alpha = x/z. \tag{17}$$

Therefore, we can write two solutions to Eq. (15.2):

$$\sin \alpha = +\frac{x}{\sqrt{x^2 + z^2}}, \qquad \cos \alpha = +\frac{z}{\sqrt{x^2 + z^2}}, \tag{18}$$

and

$$\sin \alpha = -\frac{x}{\sqrt{x^2 + z^2}}, \qquad \cos \alpha = -\frac{z}{\sqrt{x^2 + z^2}}. \tag{19}$$

In Eqs. (18) and (19), and in what follows, we take $\sqrt{x^2 + z^2} \geq 0$ (that is, we consider positive roots only). Substituting the first solution in Eq.(15.1) we get

$$x^2 + 2zR + (R^2 - H^2) - 2R\sqrt{x^2 + z^2} = 0. \tag{20}$$

The left side of Eq. (20) can be factored leading to

$$\left(z + R - H - \sqrt{x^2 + z^2}\right)\left(-z + R + H - \sqrt{x^2 + z^2}\right) = 0. \tag{21}$$

If we now substitute Eq. (19) in Eq. (15.1) we obtain

$$x^2 + 2zR + (R^2 - H^2) + 2R\sqrt{x^2 + z^2} = 0. \tag{22}$$

This equation can also be factored resulting in

$$\left(z - R - H - \sqrt{x^2 + z^2}\right)\left(-z - R + H - \sqrt{x^2 + z^2}\right) = 0. \tag{23}$$

So, one can write the solutions to Eqs. (15.1) and (15.2) as

$$\sqrt{x^2 + z^2} = z + R - H, \tag{24.1}$$
$$\sqrt{x^2 + z^2} = -z + R + H, \tag{24.2}$$
$$\sqrt{x^2 + z^2} = z - R - H, \tag{24.3}$$
$$\sqrt{x^2 + z^2} = -z - R + H. \tag{24.4}$$

Taking into account the known restrictions $\sqrt{x^2 + z^2} \geq 0$ and $z \leq H$, we conclude that the solution (24.3) is not physically acceptable and thus has to be omitted. We now apply the remaining solutions to the two different cases: $H > 0$ and $H < 0$.

**2.1. The case $H > 0$.**

In this case there are two possibilities: either $R > |H|$ or $R < |H|$. It is easily seen that when $R > |H|$, then only the solutions (24.1) and (24.2) are suitable and they describe the envelope for the family of parabolic segments.

Recall that by definition a parabola is such a curve that any point on it is at an equal distance from a given point (focus) and a given straight line (directrix). Then it becomes clear that Eqs. (24.1, 24.2) describe two parabolas. The first parabola (Eq. (24.1)) opens up and has the focal length $f_u = {}^1\!/_2\,(R - |H|)$ while the second parabola (Eq. (24.2)) opens down and has the focal length $f_d = {}^1\!/_2\,(R + |H|)$. Both parabolas have their foci at the origin (focus of the parabolic mirror). As an example, Fig.5 shows the path of the very same particle after 4 and 50 collisions.

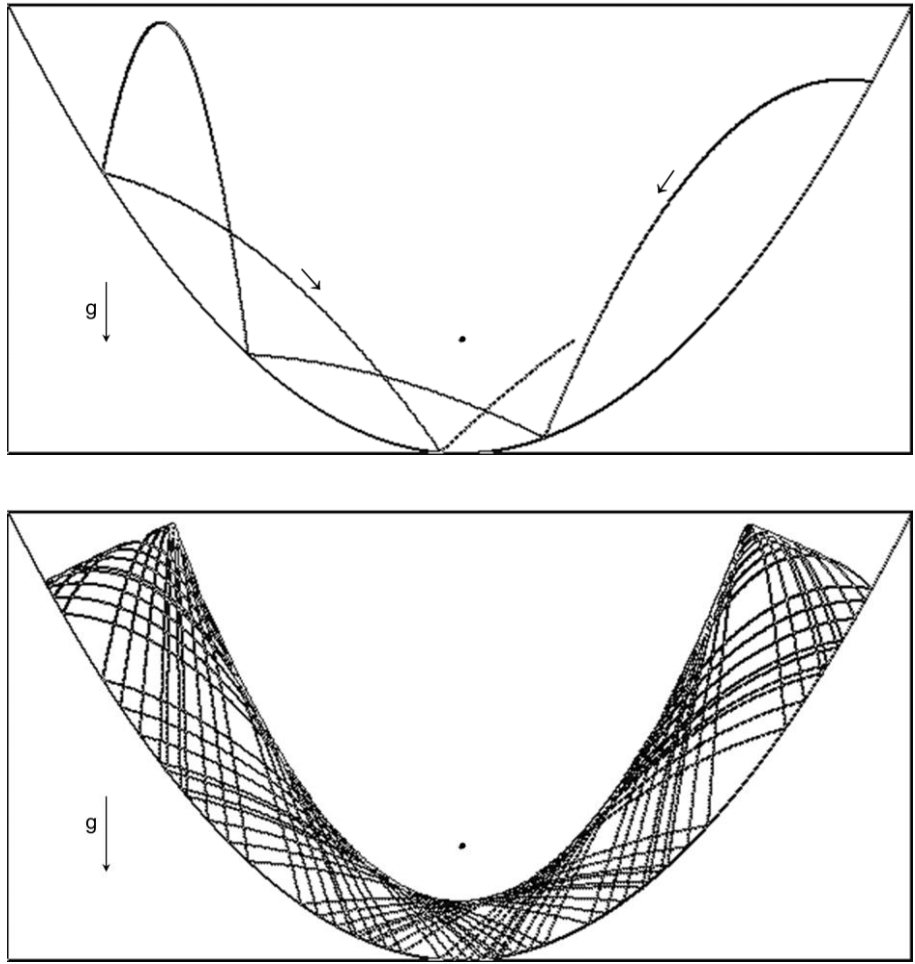

Fig.5. The case $H > 0$ and $R > |H|$. The computer-simulated path after 4 and 50 collisions

If $R < |H|$, then Eqs. (24.2) and (24.4) describe the envelope that consists of two parabolas (both open down) with $f_{d1} = {}^1\!/_2\,(R + |H|)$ and $f_{d2} = {}^1\!/_2\,(|H| - R)$. Fig.6 illustrates this case where again the trajectory of the very same particle is shown after 4 and 50 collisions.

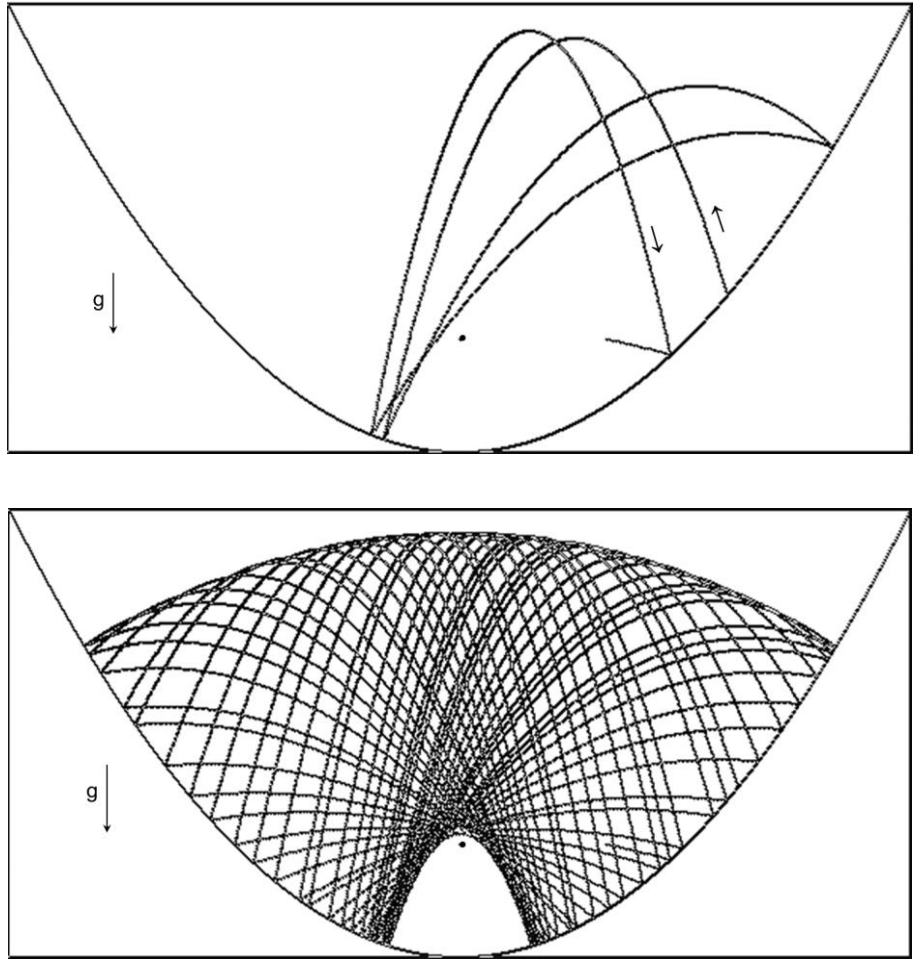

Fig.6. The case $H > 0$ and $R < |H|$. The computer-simulated path after 4 and 50 collisions

## 2.2. The case $H < 0$.

In this case there is only one possibility: $R > |H|$. Here Eqs. (24.1) and (24.2) describe an envelope for the family of parabolic arcs. This envelope consist of two parabolas (up and down) with $f_u = {}^1\!/_2\,(R + |H|)$ and $f_d = {}^1\!/_2\,(R - |H|)$. These results agree well with the results obtained earlier by Wallis et al [5] who solved a classical mechanics problem by finding the constants of motion in a three-dimensional case.

It is worth noting that, in contrast to the case $H > 0$ and $R > |H|$, the focal length of the parabola that opens down is less than the focal length of the parabola that opens up. Hence, the area bounded by the envelope is more compact. This fact is of great practical importance for designing a storage cavity for trapped cold atoms and we now estimate the maximum radius of the area on the mirror in which lie all collision points of an infinite trajectory. Obviously, this area is bounded by the line of intersection of the mirror surface ($\sqrt{x^2 + z^2} = z + 2F$) and the envelope ($\sqrt{x^2 + z^2} = -z + R - |H|$). The coordinates of the intersection points are

$$x = \pm\sqrt{2F(R - |H|)}, \tag{25}$$

$$z = {}^1\!/_2\,(R - |H| - 2F). \tag{26}$$

Let us consider a cloud of ultracold particles dropped at a height $h$ above the mirror and take an arbitrary particle in the cloud. Suppose the particle starts its trajectory at the point D (see Fig.7) with coordinates $x_i = \delta$, $z_i = -F + h$ and with initial velocity $v$. The direction of that velocity is a variable parameter. It follows from Eq. (25) that the maximum $x$-displacement of the trajectory will take place when the difference $(R - |H|)$ reaches its maximum. This difference can be found with the help of the geometric construction shown in Fig.7.

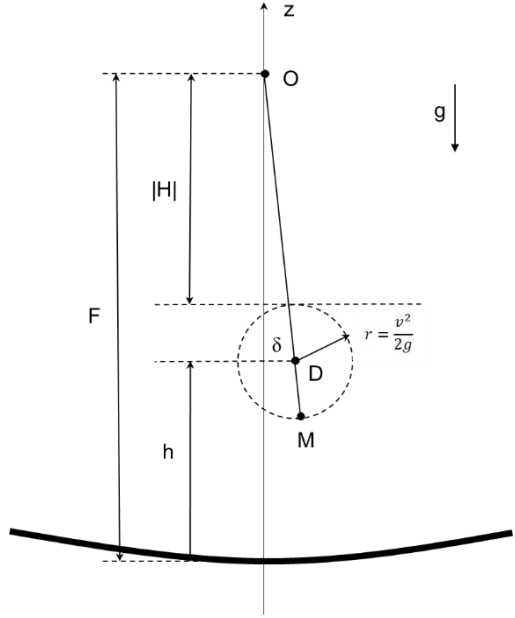

Fig.7. The determination of the maximum difference $(R - |H|)$ for a particle with velocity $v$.

It follows from Eqs. (7,8) that the focus of the first parabolic segment of the trajectory lies on a circle of radius $r = v^2/2g$ around the point $D$ and the directrix lies at a height $v^2/2g$ above $D$. Since the foci of all subsequent parabolic segments are at the same distance from the focus of the mirror $O$, we only need to find the maximum distance from O to the focus of the first parabolic segment. It can be clearly seen in Fig.7 that the maximum distance will be reached at the point $M$ where the circle $r$ is crossed by the line that goes through $O$ and $D$. Hence, the maximum distance is given by

$$R_{max} = \sqrt{(F-h)^2 + \delta^2} + \frac{v^2}{2g}. \qquad (27)$$

Since the directix is a common line for all subsequent parabolic segments, then it follows that

$$|H| = F - h - \frac{v^2}{2g}. \qquad (28)$$

In the case when $\delta \ll F$, the substitution of Eqs. (27) and (28) in Eq. (25) gives

$$x_{max} = \sqrt{2F\left(\frac{v^2}{g} + \frac{\delta^2}{2(F-h)}\right)}, \qquad (29)$$

which is in excellent agreement with [5,6].

**2.3. Construction of the collision points**

We shall now show how the geometric approach gives us another useful tool for building an infinite billiard trajectory. Suppose we know an initial point on the mirror and velocity vector at this point. In that case we know three parameters of the problem: $H$, $R$ and the focal length of the mirror $F$. It turns out that we can easily find all successive collision points by a simple geometric procedure. Fig.8 illustrates the needed geometric constructions.

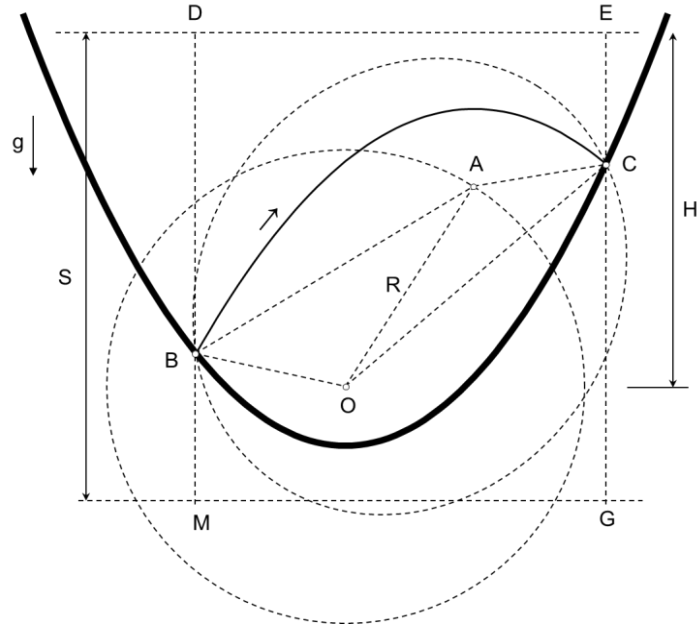

Fig.8. The geometric constructions that clarify the properties of a flight parabola

Let BC be a parabolic segment of the trajectory (we have chosen the case $H > 0$ and $R < |H|$, but the same result can be obtained for $R > |H|$ and for the case $H < 0$ as well). The point O denotes the focus of the mirror and the point A – the focus of the parabolic segment. Two horizontal dot lines depict the directrices of the mirror and flight parabola. We draw vertical lines through B and C and mark the cross points with the two directrices as D, E, M and G. By the properties of a parabola, we write $BA = BD$, $BO = BM$, $CA = CE$ and $CO = CG$. It is easy to verify that

$$BO + BA = BM + BD \equiv S, \qquad (30.1)$$
$$CO + CA = CE + CG \equiv S. \qquad (30.2)$$

The sum $S$ is the distance between the two directrices and hence it is a constant for all parabolic segments of a given trajectory. Therefore, it follows that the points B and C lie on an ellipse with its foci at O and A. We now recall that the foci of the parabolic arcs prior and after reflection are symmetric about the line that goes through the focus of the mirror and the point of collision (line OC in Fig.8). Hence, for the parabolic arc after collision we get the same ellipse but reflected in the line OC. This procedure is shown in Fig.9 for the case $R > |H|$.

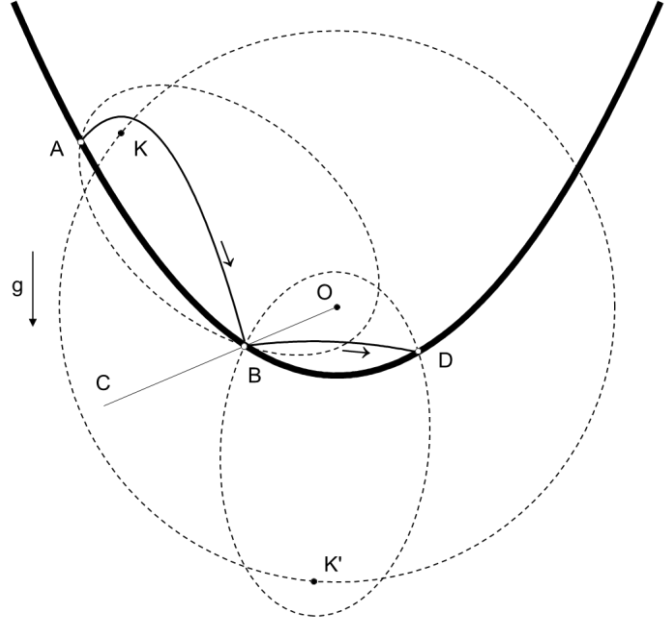

Fig.9. The procedure for building a trajectory by successive reflections of a virtual ellipse

Here AB is the first parabolic arc with the focus at K. The ellipse with its foci at O and K after reflection in the line OC (which goes through B) will be transformed into the ellipse with the foci at O and K'. The point of intersection of the new ellipse with the mirror (the point D in the figure) gives us the next collision point of the trajectory. Repeating this procedure continuously, we can find all collision points on the mirror.

From this, as a corollary, follows: If the endpoints of the flight parabola of a particle coincide with the endpoints of the focal chord of the parabolic mirror, then these two points are the only points of collision for a given particle and a billiard path is periodic, i.e. repeats itself over and over again. It is easy to verify that this happens if

$$H^2 - R^2 = 4F^2, \qquad (31)$$

where H is the z-coordinate of the directrix of parabolic arcs in the frame of reference with the origin at the focus of the mirror, $R$ is the radius of the focal circle and $F$ is the focal length of the mirror.

It should be mentioned that in computer simulations we observed either periodic or chaotic trajectories depending on an initial state of the particle. In the present paper we do not conduct any study on this subject and the interested reader may refer to [6,10,11].

## 3. Billiards in a right angle mirror

We now turn to a billiard table with an open top and with a right-angle mirror at the bottom. Here, a particle bouncing in the right angle will only reflect from plane mirrors. We first consider the properties of flight parabolas of a particle incident on and reflected from one mirror (Fig.10).

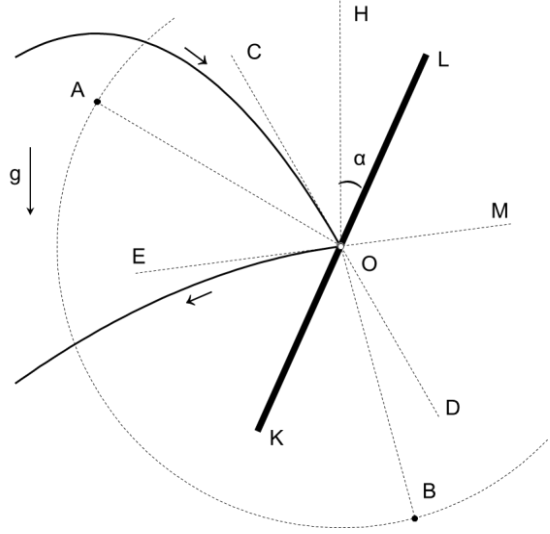

Fig.10. Reflection of a particle by a plane mirror

We shall again use the result that the focus A of the parabola made by the particle prior to collision and the focus B of the parabola made by the particle after collision are at an equal distance from the point of collision O. We draw the tangent to the first parabola (line CD) and the tangent to the second parabola (line EM) at O. We assume that the plane mirror KL makes an angle α with the vertical direction. Given the angle ∠HOA between the vertical direction OH and the line OA, we find the angle ∠HOB that the line OB makes with the vertical line OH.

By using the properties of a parabola, we obtain ∠AOC = ∠COH and ∠BOM = ∠MOH. Since we are considering only an elastic scattering (i.e. specular reflection), then the equality ∠KOE = ∠COL holds as well. After simple calculations we arrive at

$$\angle \text{HOB} = 2\pi + 4\alpha - \angle \text{HOA}, \tag{32}$$

where a positive angle is measured from the line OH in the counter-clockwise direction (e.g., the angle α in Fig.10 is negative). Eq. (32) implies that at any collision the sum (∠HOB + ∠HOA) remains constant and depends only on an inclination of the mirror, but not on an initial velocity and direction of a particle.

If $\alpha = -\pi/4$ we obtain

$$\angle \text{HOB} + \angle \text{HOA} = \pi. \tag{33}$$

Since AO = BO, it follows that in this case the points A and B lie on a common vertical line. For the next collision we again obtain the same result and so on. Hence all successive parabolic segments (flight parabolas) of the trajectory of a particle bouncing on the mirror will have their foci on a common vertical line, which we call a focal line. The position of this line depends on initial coordinates and a velocity vector of the particle. One example of such a trajectory is shown in Fig.11.

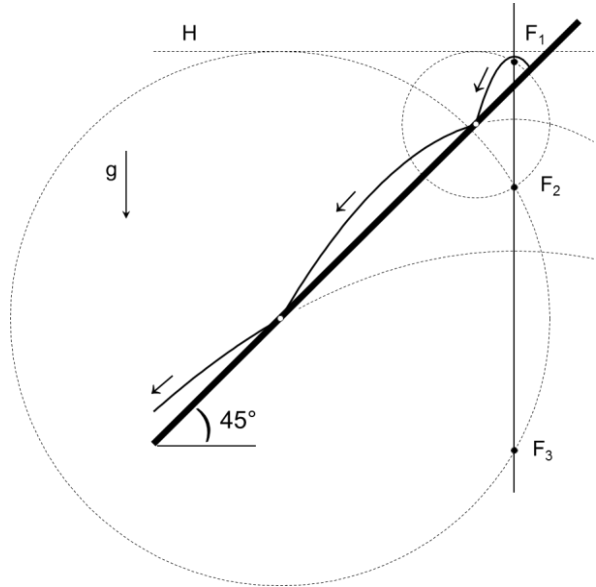

Fig.11. The path of a particle bouncing on the mirror with an inclination angle of *π/4*. The foci $F_1$, $F_2$ and $F_3$ lie on a common vertical line.

We now consider a particle bouncing on a right angle billiard table where two mirrors make angles + 45° and - 45° with the vertical direction. It is clear that the properties of foci of parabolic arcs are valid for both sides of the angle. If we now take the parabolic arc that goes from one side of the angle to the other side, then for this parabola the focus will lie on the focal line of the left mirror and on the focal line of the right mirror as well. Therefore, there is only one common focal line for both mirrors.

We have earlier seen that all parabolic segments of a given trajectory have a common directix and hence for any trajectory there are two fixed lines: a focal line and directrix. We shall use these results to get the equation of an envelope. One example of a flight parabola in a right angle billiard system is shown in Fig.12.

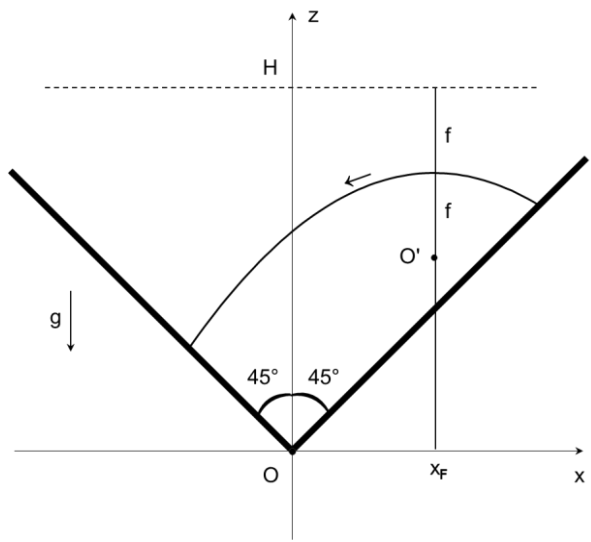

Fig.12. The flight parabola between two sides of the right angle mirror.

The point $O'$ marks the focus of the arc and *f* denotes the focal length. For convenience, we repeat here the equation (11) for that parabola in the frame of reference with the origin at $O'$:

$$-z' = \frac{x'^2}{4f} - f. \qquad (34)$$

The transformation to the coordinate system with the origin at O (vertex of the angle) is given by

$$x = x_F + x', \qquad (35.1)$$
$$z = H - 2f + z', \qquad (35.2)$$

where $x_F$ is the x-coordinate of the focal line and $H$ denotes the z-coordinate of the directix. Substituting Eq. (35.1) and (35.2) in Eq. (34) we obtain the equation

$$F(x,z,f,H) \equiv z + f - H + \frac{(x - x_F)^2}{4f} = 0, \qquad (36)$$

which describes the parabolic arc in the reference frame with the origin at O. By analogy with Eq. (14) we can say that Eq. (36) describes the family of all parabolic arcs with given $x_F$ and $H$. Now the parameter $f$ is a variable parameter within the family and determines any particular arc. Consequently, we find the envelope by solving the system

$$F(x,z,f,H) = 0, \qquad (37.1)$$
$$\partial F(x,z,f,H)/\partial f = 0. \qquad (37.2)$$

The solution of Eq. (37.2) is

$$x - x_F = \pm 2f. \qquad (38)$$

Substituting $x - x_F = +2f$ in Eq. (37.1) we obtain

$$z - H = -(x - x_F). \qquad (39)$$

This equation descries a straight line that goes through the point $(x_F, H)$ and makes an angle $+\pi/4$ with the vertical direction (recall that a positive angle is measured from the vertical line in the counter-clockwise direction).
If we put the solution $x - x_F = -2f$ in Eq. (37.1), then we get

$$z - H = x - x_F, \qquad (40)$$

which is the equation of a straight line that goes again through the point $(x_F, H)$, but this time makes an angle $-\pi/4$ with the vertical direction. Thus the envelope consists of two straight lines which go through the point $(x_F, H)$ and they are perpendicular to the sides of the right angle billiard table (see Fig.13).

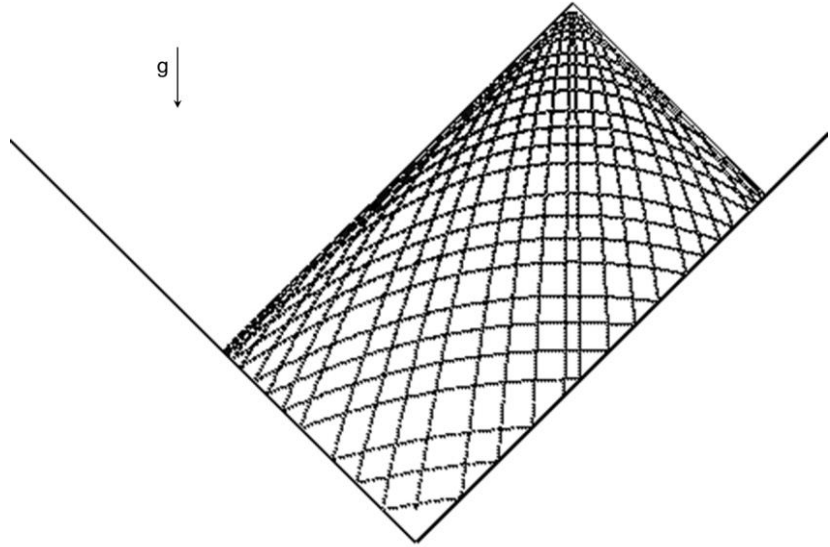

Fig.13. The computer-simulated path of a particle bouncing in the right angle billiards.

It should be pointed out that this result can also be obtained from a simple mechanical consideration. Indeed, the motion of a particle in right angle billiards can be thought of as a sum of two independent motions with velocities parallel to the sides of the billiard table [10,11]. Since these sides make a right angle and because of a specular reflection, each velocity changes independently and may be affected only by the force component parallel to the associated side of the billiard table. Hence, each separate motion represents a simple bouncing of a particle in a homogeneous force field, where as a force we take the corresponding component of the full force. The height of bouncing in either direction remains constant for each independent motion and so the area available for the particle is bounded by two straight lines normal to the sides of the right angle.

**Conclusion**

The geometric approach was shown to be a useful tool in studying billiards in a homogeneous force field. In the present paper we applied this method to the two specific tables and studied the geometric properties of the billiard trajectories. We have shown that any particular infinite trajectory can be characterized by two lines common for all sequential segments of that trajectory: a focal circle and directrix for parabolic billiards; and a focal line and directrix for right angle billiards. With this result we have found the equation of an envelope for every trajectory. We have also derived the condition for getting a simple cyclic trajectory. However, we did not study the problem of general requirements for chaotic and cyclic trajectories since this was not a subject of our paper.

In addition, we would like to report some characteristic features of the trajectories observed in the considered billiard systems. First, in right angle billiards every parabolic arc of the trajectory makes a right angle with the vertical line that goes through the cusp on the envelope (see Fig. 13). This follows from Eq. (33) and comments there. As the next example, we consider a trajectory of the type shown in Fig. 5. Let us draw a vertical line through the cusp on the envelope and look at the points of intersection of this line with the parabolic arcs of the trajectory. One can see that tangents to the arcs at these points go through the focus of the mirror. Although this observed feature was left without proof, we do believe it is worthy of notice because of a potential practical application.

We have presented the geometric approach applied only to two billiard tables with specific boundaries, but it seems to be important to try to apply this method to other billiard tables as well.